\documentclass[twocolumn]{revtex4}
\usepackage{amsmath}
\usepackage{amsfonts}
\usepackage{graphicx}
\usepackage{setspace}
\usepackage{bm}
\newcommand{\bff}{ \mathbf{f}}

\begin{document}

\title{Possibility and Impossibility of the Entropy Balance in Lattice
Boltzmann Collisions}
\author{Alexander N. Gorban and Dave Packwood}
\affiliation{Department of Mathematics, University of Leicester,
United Kingdom}

\begin{abstract}
We demonstrate that in the space of distributions operated on by
lattice Boltzmann methods that there exists a vicinity of the
equilibrium where collisions with entropy balance are possible and,
at the same time, there exist an area of nonequilibrium
distributions where such collisions are impossible. We calculate and
graphically represent these areas for some simple entropic
equilibria using single relaxation time models. Therefore it is
shown that the definition of an entropic LBM is incomplete without a
strategy to deal with certain highly nonequilibrium states. Such
strategies should be explicitly stated as they may result in the
production of additional entropy.
\end{abstract}

\maketitle

\section{Introduction}

Lattice Boltzmann schemes are a type of discrete algorithm which can
be used to simulate fluid dynamics and more \cite{Succi,Benzi}.
Although such a method can be derived as a discretization of the
fully continuous Boltzmann equation, some thermodynamics properties
may be lost in this process. The Entropic lattice Boltzmann method
(ELBM) was invented first in 1998 as a tool for the construction of
single relaxation time lattice Boltzman models which respects a
$H$-theorem \cite{Htheorem,SucciRevModPhys}. For this purpose,
instead of the mirror image with a local equilibrium as the
reflection center, the entropic involution was proposed, which
preserves the entropy value. Later, it was called the
\textit{Karlin-Succi involution} \cite{gorban06}.

Nevertheless, controlling the proper entropy balance remained until
recently a challenging problem for many lattice Boltzmann models
\cite{Nonexist}. Some discussions of  modern ELBM implementations
and results were published recently \cite{KarlinSucciComment}.

The distribution functions at the centre of lattice Boltzmann
methods are often referred to and understood as particle
densities. Of course for such an interpretation to be
meaningful the distribution function should be strictly
positive. Despite this some lattice Boltzmann implementations
may, as a numerical scheme, tolerate negative population
values. An ELBM usually involves an evaluation of a Boltzmann
type entropy function, which does not exist for negative
populations, hence such an ELBM cannot ever tolerate a negative
population value. Due to this there are population values for
which an entropic involution cannot be performed. A complete
definition of an ELBM must include a strategy for what to do in
such a situation. The choice of such a strategy should be
explicitly given in any definition of an ELBM as it may have
side-effects with modification of dissipation which should be
understood separately from the influence of the proper entropy
balance.

In this paper we study the regions in the spaces of
distributions (populations) where collisions with
entropy preservation are possible (near the equilibrium) and
where they are impossible (sufficiently far from the
equilibrium) and demonstrate that both such areas always exist
apart some trivial degenerated cases.

\section{Single Relaxation Time LB Schemes}

\label{sec:single} For fluids, LB systems can be derived as a discretization
of the Boltzmann Equation
\begin{equation}
{\partial}_{t} f + \mathbf{v} \cdot {\partial}_{ \mathbf{x}} f =
Q(f)
\end{equation}
where $f \equiv f( \mathbf{x}, \mathbf{v},t)$ is a one particle distribution
function over space, velocity space and time and $Q(f)$ represents the
interaction between particles, sometimes called a collision operation. A
particular example of the interaction $Q(f)$ is the Bhatnagar-Gross-Krook
equation
\begin{equation}
Q(f) = -\frac{1}{\tau}(f - f^{\mathrm{eq}}).
\end{equation}
The BGK operation represents a relaxation towards the local
equilibrium $f^{ \mathrm{eq}}$ with rate $1/\tau$. The
distribution $f^{\mathrm{eq}}$ is given by the Maxwell
Boltzmann distribution,
\begin{equation}
f^{\mathrm{eq}} = \frac{\rho}{(2\pi T)^{D/2}}\exp\left(\frac{-( \mathbf{v} -
\mathbf{u})^2}{2T}\right).
\end{equation}
The macroscopic quantities are available as integrals over velocity space of
the distribution function,
\begin{equation*}
\rho  = \int f \; \mathrm{d} \mathbf{v}, \; \rho \mathbf{u}  = \int
\mathbf{v} f \; \mathrm{d} \mathbf{v}, \;\rho \mathbf{u}^2 + \rho T
= \int \mathbf{v}^2 f \; \mathrm{d} \mathbf{v}.
\end{equation*}

A discrete approximation to these integrals is the first ingredient
to discretize this system. The scalar field of the population
function (over space, vector space and time) becomes a sequence of
vector fields (over space) in time $f_i( \mathbf{x},n_t \;\epsilon),
n_t \in \mathbb{Z}$, where the elements of the vector each
correspond with an element of the quadrature. Explicitly the
macroscopic moments are given by,
\begin{equation*}
\rho  = \sum_{i=1}^n f_i ,\; \rho \mathbf{u} = \sum_{i=1}^n
\mathbf{v}_i f_i , \; \rho \mathbf{u}^2 + \rho T  = \sum_{i=1}^n
\mathbf{v}_i^2 f_i .
\end{equation*}
The complete discrete scheme is given by
\begin{equation}
f_i( \mathbf{x} + \epsilon \mathbf{v}_i,t+\epsilon) = f_i( \mathbf{x},t) +
\omega (f^{\mathrm{eq}}_i( \mathbf{x},t) - f_i( \mathbf{x},t) )
\label{eq:algorithmUnchained}
\end{equation}
where $\epsilon$ is the time step. For this system a discrete
equilibrium must be used. The choice of the velocity set $\{
\mathbf{v}_1 , \ldots ,  \mathbf{v}_n \}$ and the discrete
equilibrium distribution $f^{\mathrm{eq}}_i$ should provide the best
approximation of the transport equations for the moments by the
discrete scheme (\ref{eq:algorithmUnchained}).

\section{ELBM}
\label{sec:ELBM}

In the continuous case the Maxwellian distribution maximizes
entropy, as measured by the Boltzmann $H$ function, and therefore
also has zero entropy production. In the context of lattice
Boltzmann methods a discrete form of the $H$-theorem has been
suggested as a way to introduce thermodynamic control to the system
\cite{Htheorem,Boghosian}.

A variation on the LBGK is the ELBGK \cite{ELBM}. In this family of
methods, the equilibria are defined as the conditional entropy
maximizers under given values of macroscopic variables ({\em
entropic equilibria}). The entropies have been constructed in a
lattice dependent fashion in \cite{LatticeEntropies}. A slightly
different notation is used for the lattice Boltzmann algorithm,
\begin{equation}
f_i( \mathbf{x} + \epsilon \mathbf{v}_i,t+\epsilon) = f_i( \mathbf{x},t) +
\alpha \beta (f^{\mathrm{eq}}_i( \mathbf{x},t) - f_i( \mathbf{x},t) ).
\label{eq:alphabeta}
\end{equation}
The single parameter $\omega$ is replaced by a composite parameter
$\alpha \beta$. In this case $\beta$ controls the viscosity and
$\alpha$ is varied to ensure a constant entropy condition according
to the discrete $H$-theorem. With knowledge of the entropy function
$S$, $\alpha$ is found as the non-trivial root of the equation
\begin{equation}
S(\mathbf{f}) = S(\mathbf{f} + \alpha(\mathbf{f}^\ast -
\mathbf{f})). \label{eq:ELBMCondition}
\end{equation}
The trivial root $\alpha = 0$ returns the entropy value of the
original populations. ELBGK then finds the non-trivial $\alpha$ such
that (\ref{eq:ELBMCondition}) holds. This version of the BGK
collision one calls entropic BGK (or EBGK) collision. A solution of
(\ref{eq:ELBMCondition}) must be found at every time step and
lattice site. The EBGK collision obviously respects the Second Law
(if $\beta \leq 1$), and simple analysis of entropy dissipation
gives the proper evaluation of viscosity.

In general the entropy function is based upon the lattice. For
example, in the case of the simple one dimensional lattice with
velocities $ \mathbf{v} = (-c,0,c)$ and corresponding populations $
\mathbf{f} = (f_-,f_0,f_+)$ an explicit Boltzmann style entropy
function is known \cite{LatticeEntropies}:
\begin{equation}
S(\mathbf{f}) = -f_-\log(f_-) - f_0\log(f_0/4) - f_+\log(f_+).
\label{eq:ELBMEntropy}
\end{equation}

\section{Regions of Existence and Non-existence of Entropic Involution}

Let us study the entropic involution in the distribution simplex
$\Sigma$ given by $\sum f_i =const>0$, $f_i \geq 0$.

\begin{figure*}
\includegraphics[width=0.9\textwidth]{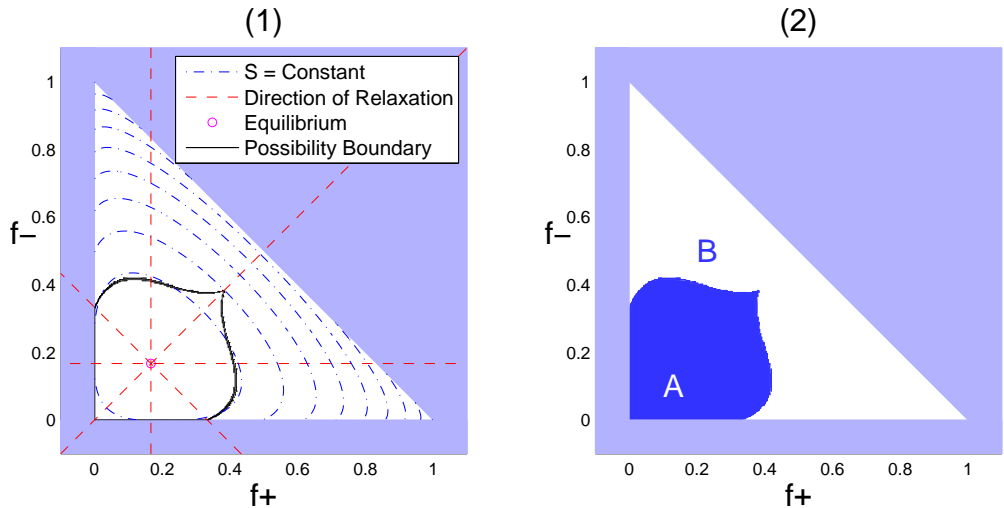}
\caption{The simplex $\Sigma$ is given by the white background. (1)
Populations relax through the equilibrium given by the single point
to an equal entropy point, if possible. The boundary of this
possibility is given. (2) The regions $A$  (the entropic involution
is possible) and $B$ (the involution is impossible) as subsets of
the simplex divided by this boundary are presented.
\label{fig:onlyrho}}
\end{figure*}
\begin{figure*}
\includegraphics[width=0.9\textwidth]{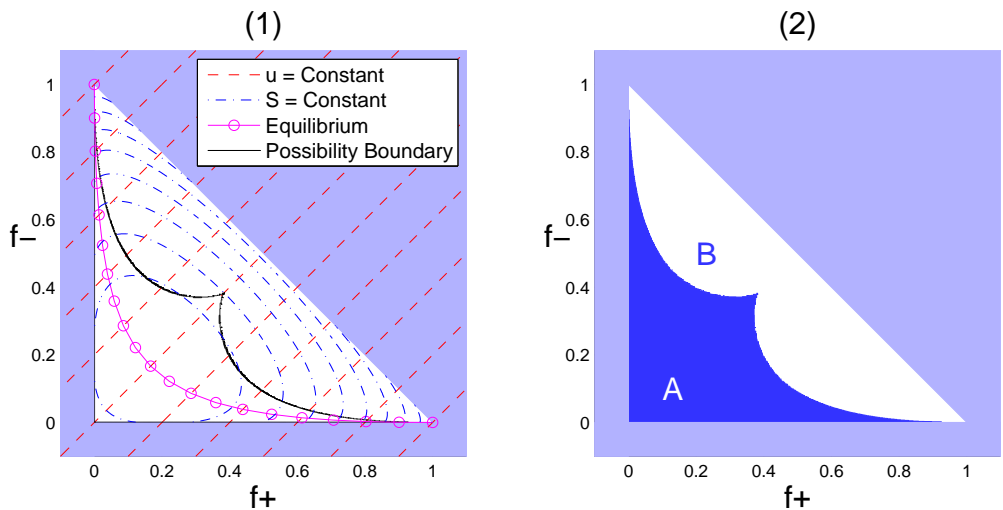}
\caption{The simplex $\Sigma$ is given by the white background. (1)
Populations relax through the their corresponding equilibrium point
along the line given by constant $u$ to an equal entropy point, if
possible. The boundary of this possibility is given. (2) The regions
$A$ and $B$  separated by this boundary are presented.
\label{fig:withu}}
\end{figure*}

Let us prove that under very natural assumptions about some properties of the entropy
that the simplex of distributions can be split into two
subsets $A$ and $B$: in the set $A$ the entropic involution exists,
and for distributions from the set $B$ equation
(\ref{eq:ELBMCondition}) has no non-trivial solutions. Both sets $A$
and $B$ have non-empty interior (apart of a trivial symmetric
degenerated case).

Let the entropy $S$ be a strictly concave continuous function in the
distribution simplex $\Sigma$. We assume also that $S$ is twice
differentiable,  the Hessian of $S$, $\partial^2 S/
\partial f_i\partial f_j$, is negative definite in the interior of the
simplex, $\Sigma_+$, where $\sum f_i =const$, $f_i > 0$ and the
global maximizer of $S$, the equilibrium, belongs to the
interior of the simplex.

For example, the relative Boltzmann entropy, $S=-\sum f_i (\ln
(f_i/W_i)-1)$, $W_i >0$, satisfies these conditions, because $f\ln f
\to 0 $ when $f\to 0$ and $\partial^2 S/
\partial f_i\partial f_j= - \delta_{ij}/f_i$, whereas the relative {\it Burg
entropy} $S=\sum W_i (\ln (f_i/W_i))$ does not satisfy these
conditions because it does not exist on the border of the simplex.

Macroscopic variables  are linear functions of $\bff$.
The sets with given values of the macroscopic variables in the
simplex $\Sigma$ are polyhedra, intersections of $\Sigma$ with
linear manifolds with the given values of moments. We assume
that in any such a polyhedron the entropy achieves its
(conditionally) global maximum at an internal point. This
assumption holds for the Boltzmann relative entropy because of the
logarithmic singularity of the ``chemical potentials" $\mu_i =
\ln (f_i/W_i)$ on the border of positivity. These maximizers
are equilibria. If $\bff$ is sufficiently close to a
positive equilibrium then, due to the implicit function
theorem, the nontrivial solution to equation
(\ref{eq:ELBMCondition}) exists and it gives $\alpha=2 +
o(\mathbf{f}-\mathbf{f}^*)$. The value $\alpha=2$ corresponds
to the mirror image, the small term
$o(\mathbf{f}-\mathbf{f}^*)$ gives the corrections to the value
$\alpha=2$. Therefore, in some vicinity of the equilibrium the
entropic involution exists.

To prove the existence of the area where entropic involution is
impossible, let us consider one polyhedron with given values of
the macroscopic variables and a positive equilibrium. The local
minima of the entropy in this polyhedron are situated at the
vertices. At least one of them is a global minimum. Let this
vertex be $\bff^\mathbf{v}$. Let us draw a straight line $l$ through
points $\bff^\mathbf{v}$ and $\mathbf{f}^*$. The intersection $l \cap
\Sigma$ is an interval and $S$ achieves its global minimum on
this interval at the point $\bff^\mathbf{v}$. If the dimension of the
polyhedron is more than one then the opposite end of this
interval is not even a local minimum of $S$ in the polyhedron
and the entropic involution does not exists for  $\bff^\mathbf{v}$
and some vicinity around it.

A special degeneration is possible when the polyhedra are
one-dimensional, i.e. intervals, and the values of the entropy
at both ends of each interval coincide. For example, for
two-dimensional distributions, $f_+, f_-$, the entropy $S==f_+
\ln f_+-f_-\ln f_-$ and the macroscopic variable $\rho=f_+ +
f_-$. Apart from such symmetric one-dimensional cases there exists
an area near the maximally non-equilibrium vertex $\bff^\mathbf{v}$
where the entropic involution cannot be defined. Such an area
may also exist near some other vertices, where local entropy
minima are reached.

For the Burg entropy, the entropic involution is always possible
\cite{Boghosian} because it tends to $-\infty$ at the border of
positivity. The same is true, for the relative entropy of the form
$S=-\beta^{-1}\sum W_i ((W_i/f_i)^{\beta}-1)$ that tends to the Burg
entropy when $\beta\to 0$ \cite{GoGoJudge2011}. This {\it negative
brunch} of the relative {\it Tsallis entropy} is less known. The
standard Tsallis entropy \cite{Tsallis} is finite at the border of
positivity, hence, collisions with entropy preservation are not
always possible for it.

We now demonstrate the population function values where the
involution cannot be performed for some simple examples. We use the
standard 1-D lattice described in Section \ref{sec:ELBM} with the
discrete equilibrium given in Eq \ref{eq:ELBMEntropy}. We begin with
an LBM with only one conserved moment in collision, namely density.
The equilibrium is  $f_-^* = \frac{\rho}{6}, \; \; f_0^* =
\frac{2\rho}{3}, \; \; f_+^* = \frac{\rho}{6}.$

In Fig.~\ref{fig:onlyrho}, the simplex $\Sigma$ of positive
populations with a fixed density $\rho=1$ is the triangle given by
the intersection of three half-planes,  $f^+ > 0, \,f^- >0$, and $1
- f^+ - f^- > 0$. Within that region we plot several entropy level
contours $S( \mathbf{f}) = c$ and the unique equilibrium point. The
region is divided into the parts where the entropic involution is
possible (around the equilibrium) and where it is impossible.

A more common use of lattice Boltzmann involves a second fixed
moment, momentum. The entropic equilibria used by the ELBGK are
available explicitly as the maximum of the entropy function
(\ref{eq:ELBMEntropy}),
\begin{equation*}
f_{\mp}^* = \frac{\rho}{6}(\mp 3u - 1 + 2\sqrt{1\!+\!3u^2}), \;
f_0^* = \frac{2\rho}{3}(2 - \sqrt{1\!+\!3u^2}).
\end{equation*}
In this case the dimension of the equilibrium is one greater. In
Fig.~\ref{fig:withu} all relaxation occurs parallel to the lines of
constant $u$. The region where entropic involution is possible is
again given.

In each experiment the region is discretized into many individual
points. For each point a value for $\alpha$ is attempted to be
found. The method used is simply to begin with a guess of $\alpha=1$
and then add increments of $10^{-3}$ until a solution of Eq.
\ref{eq:ELBMCondition} occurs, or the edge of the positivity domain
is reached. This method would be inappropriate to use in a usual
ELBM, due to the very large computational cost, but it is very
robust and hence useful for this experiment with many higly non-equilibrium distributions. Another approach (with
the same result) implies calculation of the entropic involution for
all the boundary points where it exists. In this method we draw a
straight line $l$ through a boundary point $\mathbf{f}$ and the
equilibrium and find the intersection $l \cap \Sigma$ which consists
of all points on $l$ with non-negative coordinates. One end of this
interval is $\mathbf{f}$, another end is also a boundary point,
$\mathbf{f}'$. The entropic involution for $\mathbf{f}$ exits if and
only if $S(\mathbf{f})'\leq S(\mathbf{f})$. After we check this
inequality, we can solve Eq. (\ref{eq:ELBMCondition}). The
images of these involutions form the border that separates sets $A$
and $B$ (see Figs).

\section{Conclusion}

The entropic involution is not always possible to perform. We
have demonstrated that apart some special one-dimensional
spaces of distributions with additional symmetry there exist
domains where collisions with the preservation of entropy
are not possible. We illustrated this statement by some simple
and well known examples of ELBGK systems for which we directly
calculated the areas where entropic collisions exist and
where they do not exist.

Such phenomena should be observable in all ELBM schemes with the
classical entropies: there exists a vicinity of the equilibrium
where the entropic involution is possible but for some areas of
non-equilibrium distributions there exists no non-trivial root of
equation (\ref {eq:ELBMCondition}).  A collision which preserves
entropy does not exist for this area.  Therefore, for the regimes
close to equilibrium (the vicinities $A$ of equilibria, Figs 1,2),
ELBM schemes guaranty the precise balance of the entropy and for
more nonequilibrium regimes, when at some sites the distribution
belonges to sets $B$, ELBM schemes work as {\it limiters}
\cite{Limiters}. with additional dissipation. It is necessary for
any complete definition of an ELBM algorithm to prescribe what to do
when the involution is not possible. A reasonable choice would be to
over-relax the maximum amount possible while maintaining positive
population values. Such a technique is independently in use as a
stabilizer for lattice Boltzmann schemes, sometimes called the
`positivity limiter' \cite{BGJ,Limiters,Li,Servan,Tosi}. An effect
of this operation is a local increase in viscosity/entropy
production. Hence, if an ELBM were to apply such a scheme it would
necessarily break the proper entropy balance. In this sense, ELBM
belongs to a large family of add-ons that regularise LBM by the
management of the addtional dissipation \cite{Add-ons}.

\end{document}